\documentclass[twocolumn,seceq]{jpsj2}

\def\runauthor{Tatsuya Nagao and Jun-ichi Igarashi}

\title{Resonant X-Ray Scattering from URu$_{2}$Si$_{2}$}

\author{Tatsuya Nagao\thanks{E-mail address: tnagao@phys.sci.gunma-u.ac.jp}
and Jun-ichi Igarashi$^{1}$}

\inst{Fuculty of  Engineering, Gunma University, 
      Kiryu, Gunma 376-8515 \\
$^{1}$Faculty of Science, Ibaraki University, Mito, Ibaraki 310-8512}

\recdate{\today}

\abst{
Based on a localized crystal electric field model for the U$^{4+}$
in the $(5f)^2$-configuration, we analyze the resonant x-ray scattering
spectra around U $M_{\rm IV}$ and $M_{\rm V}$ edges in URu$_2$Si$_2$,
taking full Coulomb and spin-orbit interactions into account.
We consider two level schemes, a singlet model of Santini and Amoretti and
a doublet model of Ohkawa and Shimizu, and assume the antiferroquadrupolar
order and the antiferromagnetic order as candidates for the ambient pressure
phase and the high pressure phase.
It is found that the spectral shapes as a function of photon energy
are independent of the assumed level scheme, but are
quite different between the antiferroquadrupole
and antiferromagnetic phases,
This may be useful to determine the character of the ordered phase.
}

\kword{resonant x-ray scattering, URu$_2$Si$_2$, singlet scheme, 
doublet scheme}

\begin{document}
\maketitle

\section{\label{sec:1}Introduction}

\begin{figure}[t]
\begin{center}
\includegraphics[width=7.50cm]{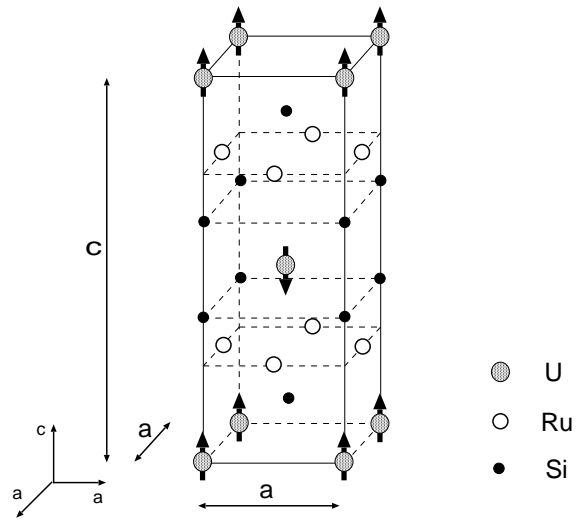}
\end{center}
\caption{\label{fig.struc} Crystal structure of body-centered 
tetragonal URu$_{2}$Si$_{2}$. The arrows show the magnetic structure 
corresponding to a wave vector ${\bf Q}=(0 0 1)$.}
\end{figure}

The ternary intermetallic compound URu$_2$Si$_2$
has been attracting much attentions
since the discovery of the coexistence of antiferromagnetic (AFM)
ordering below $T_{\rm 0}=17.5$ K and superconductivity
below $T_{\rm C}=1.2$ K.\cite{Schlabitz86} 
The ordering pattern of the AFM order is type-I  structure
with the propagation vector ${\bf Q}=(0 0 1)$ along the $c$-axis 
(Fig. \ref{fig.struc}).
The phase transition at $T_{\rm 0}$ is characterized by bulk anomalies
such as
specific heat\cite{Palstra85}, linear susceptibility,\cite{Palstra85}
non-linear susceptibility,\cite{Miyako91,Ramirez92}
thermal expansion\cite{DeVisser86} and
electrical resistivity\cite{Palstra86},
strongly indicating that the transition is of second order.

The observed magnetic moment is unusually tiny
($\mu_{\rm ord} \sim (0.03 \pm 0.01) \mu_{\rm B}$ per U ion at saturation).
\cite{Broholm87,MacLaughlin88,Fak96}
It is too small to account for the observed bulk anomalies.
The magnetic excitation spectra shows energy gap 
below $T_{\rm 0}$ according to neutron scattering experiments.\cite{Mason91}
Under the magnetic field, the magnetic moment and the magnetic excitation
gap exhibit the different field dependencies.\cite{Mason95}
Based upon these observations, it is widely believed that,
the AFM order below $T_{\rm 0}$ is merely a secondary order
and there exists some
unknown primary order, so-called "hidden order (HO)".\cite{Shah00}
As for the true nature of the HO phase, 
various competing theoretical possibilities have been demonstrated.
For instance, the antiferroquadrupole (AFQ) order,
\cite{Santini94,Santini98,Tsuruta00,Takahashi01,Ohkawa99}
the multispin correlation,\cite{Barzykin95}
the unconventional spin density wave,\cite{Ikeda98}
inhomogeneous orbital current,\cite{Chandra02.1,Chandra02.2,Chandra03}
Jahn-Teller distortions\cite{Kasuya97}
and so on were proposed. 
However, it is safe to say that the microscopic mechanism
governing the whole phenomena is still in a matter of controversy.

Recently, neutron scattering and $^{29}$Si NMR experiments were performed
under 
pressure.\cite{Amitsuka99,Amitsuka00,Matsuda01,Matsuda03,Motoyama03.1,Yokoyama02}
The measurements are interpreted that
not the magnitude of the AFM moment but the
volume fraction of the AFM phase is tiny below $T_{\rm 0}$.
The volume fraction of the AFM phase is estimated about
1 \% at ambient pressure while the magnitude of the moment is 
0.25 $\mu_{\rm B}$ per U ion.
Another discovery of importance is the presence of
the phase transition around $p_{\rm C}=1.5$ GPa,
accompanied by the discontinuous changes of the quantities 
such as $T_{\rm 0}$, 
the lattice constant and the value of the magnetic moment.
In high pressure phase, the AFM order extends uniformly throughout 
the entire sample. 
It exhibits a three dimensional AFM Ising behaviors
with the saturated staggered moment $\sim 0.4 \mu_{\rm B}$/U
and with the same magnetic structure as in the
ambient pressure (HO) phase.\cite{Amitsuka99}

One of the promising efforts to elucidate the nature of the order
parameter in the $d$ and $f$ electron systems is the
resonant x-ray scattering (RXS) measurement.
By utilizing the facts that the method is element specific and 
level specific, and that the RXS intensity generally shows
special incident photon polarization and
azimuthal angle dependencies, we can deduce valuable information to
understand the physical properties of subjects.
In $4f$ systems, for example,
the RXS peaks at the Ce $L_{\rm III}$ absorption edge 
in CeB$_{6}$ are interpreted as the direct observation of the
AFQ ordering\cite{Nakao01,Nagao01,Igarashi02,Lovesey02} while  the peaks
at the Dy $L_{\rm III}$ edge in DyB$_2$C$_{2}$ are explained as 
the consequence of lattice distortion.
\cite{Tanaka99,Hirota00,Matsumura02,Igarashi03}

For actinide compounds, the $M_{\rm IV,V}$ edge resonance
is available for superlattice Bragg spots.
Since the dipole process gives rise to a transition from the $3d$ states to
the $5f$ states which drive ordering,
the RXS signal is more direct than those at the $K$-edge 
in the transition-metal compounds.
In transition metal oxides,
the RXS signals at the $K$-edge involve the transition from the $1s$ state
to the $4p$ states. Since the $4p$ states are considerably extending
in space, they are very sensitive to electronic structures
at neighboring sites. Thereby the RXS signal on the orbital ordering 
superlattice spots are mainly controlled by lattice distortion.
\cite{Elfimov99,Benfatto99,Takahashi99}

Isaacs \textit{et al}. reported the resonant enhancements 
corresponding to the U $M_{\rm IV}$ edge
upon URu$_{2}$Si$_{2}$ at ambient pressure below $T_{\rm 0}$.
\cite{Isaacs90}
The measured RXS peak intensity at the prohibited Bragg spot $(003)$
develops linearly over an unusually wide
range of temperature from $T_{\rm 0}$ down to about 3 K,
then ceases to grow when superconductivity sets in.
The width of the spectra is evaluated about 5 eV.
Later, Lidstr\"{o}m \textit{et al}. have observed
the peak intensities at U $M_{\rm IV}$ and $M_{\rm V}$ edges
from U$_{1-x}$Np$_x$Ru$_2$Si$_2$ alloys with $x=0.1, 0.5$ and $1.0$
also using the RXMS technique.\cite{Lidstrom00}

In this paper, we analyze the RXS spectra in the low temperature phases 
of URu$_2$Si$_2$ on the basis of a localized model under 
the crystal electric field (CEF).
Although the CEF-like excitations are not clear in the high temperature phase,
a localized picture can explain well the behavior of various quantities
such as the specific heat and the magnetic susceptibility.
\cite{Santini94,Amitsuka99,Amitsuka00}
Since the CEF parameters are not known,
the level schemes are not definitely determined.
There are two prevailing schemes,
the ``singlet" scheme of Santini and Amoretti\cite{Santini94}
and the ``doublet" scheme of Ohkawa and Shimizu.\cite{Ohkawa99}
On the basis of the two level schemes, we consider 
both the AFQ and AFM phases, and calculate the RXS spectra 
as a function of photon energy and of azimuthal angle.
We find that the calculated spectral shapes are independent of 
the assumed scheme, although the intensities are quite different.
\cite{com1}
Thereby the RXS spectral shapes depend only on the assumed order parameter.
The RXS spectral shapes as a function of energy are found quite different
between the AFQ phase and the AFM phase. In principle,
this may make it possible to distinguish whether the experimental 
signals come from the HO phase or from the AFM domains.
Unfortunately available experimental spectra are not clear enough 
to distinguish those differences.\cite{Isaacs90,Lidstrom00}
Shishidou \textit{ et al}.\cite{Shishidou00}
calculated the RXS spectra in the AFQ phase within the doublet scheme.
The present calculation agrees with their result.

The present paper is organized as follows.
In \S \ref{sect.2}, we briefly summarize the theoretical framework 
of RXS.
In \S \ref{sect.3}, we discuss two level schemes under the CEF 
on the basis of the localized $5f$ model,
and the AFQ and AFM states as candidates of the ground state
under these schemes.
The intermediate states are briefly discussed.
In \S \ref{sect.4}, numerical results of the RXS spectra 
are presented in comparison with the experiments.
A summary and discussion are given in \S \ref{sect.5}.
In Appendix, general expressions of the matrix elements of the scattering
amplitude are given. It is proved that the RXS spectral shapes are 
independent of the assumed level schemes.

\section{\label{sect.2}Theoretical Framework of RXS}

\begin{figure}[t]
\begin{center}
\includegraphics[width=7.50cm]{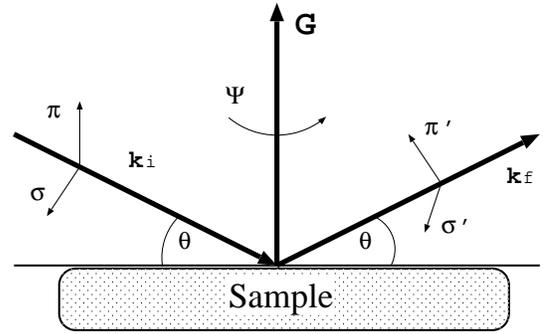}
\end{center}
\caption{\label{fig.geom}Geometry of the RXS experiment. 
Photon with polarization $\sigma$ or $\pi$
is scattered into the state of
polarization $\sigma'$ or $\pi'$ at Bragg angle $\theta$.}
\end{figure}

We consider the experimental situation of RXS shown in Fig. \ref{fig.geom}.  
The incident photon with frequency $\omega$
and wave vector ${\bf k}_{i}$ is scattered into the state with
frequency $\omega$ and wave vector ${\bf k}_{f}$.
The sample is rotated around the scattering vector 
${\bf G} (={\bf k}_{f}-{\bf k}_{i})$ by azimuthal angle $\psi$.
The resonant enhancement is found at the U $M_{\rm IV}$ and $M_{\rm V}$ 
absorption edges, where the $3d$ core electron is virtually
excited to the $5f$ states then falls to the core level 
in the dipolar ($E_1$) process.  
The tuning energy for $M_{\rm IV}$ edge is around 3.728 keV 
and that for $M_{\rm V}$ is around 3.550 keV.
The scattering tensor can be well approximated by a sum of the contributions 
from each site of the created core hole.
Therefore the RXS intensity in the $E_1$ process is given by
\begin{equation}
 I_{\mu\to\mu'} ({\bf G},\omega) \propto
 \left| \sum_{\alpha\alpha'}P'^{\mu'}_{\alpha}
    M_{\alpha\alpha'}({\bf G},\omega)P^{\mu}_{\alpha'}
  \right|^2 ,
\label{eq.rxs.intensity}
\end{equation}
with 
\begin{align}
 M_{\alpha\alpha'}({\bf G},\omega) 
& = \frac{1}{\sqrt{N}}
  \sum_j\sum_{\Lambda} {\rm e}^{-i{\bf G}\cdot{\bf r}_j} \nonumber \\
& \times    \frac{\langle\psi_0|x_\alpha(j)|\Lambda\rangle
  \langle \Lambda|x_{\alpha'}(j)|\psi_0\rangle}
       {\hbar\omega-(E_{\Lambda}-E_0)+i\Gamma},  \label{eq.dipole}
\end{align}
where $N$ is the number of U ions. The $| \psi_0 \rangle$ represents the ground state with energy $E_0$,
and $|\Lambda\rangle$ represents the intermediate state
with energy $E_{\Lambda}$.
The dipole operators $x_\alpha(j)$'s are defined as
$x_1(j)=x$, $x_2(j)=y$, and $x_3(j)=z$ in the coordinate frame fixed 
to the crystal axes with the origin located at the center of site $j$.
The $P^\mu$ and $P'^{\mu'}$ denote the geometrical factors
for the incident photon and the scattered photon,
whose explicit expressions are given in ref. 37.
The $\Gamma$ describes the life-time broadening width of the core hole.

\section{\label{sect.3}Electronic Structure} 

\subsection{The ground state}
We adopt a localized description that each U ion is 
in the $(5f)^2$-configuration.
The Hamiltonian of U ions consists of
the intra-atomic Coulomb interaction between $5f$ electrons and 
the spin-orbit interaction (SOI) of $5f$ electrons, which are represented 
by 91 states in the $(5f)^2$-configuration.
We evaluate the Slater integrals for the Coulomb interaction in the $5f$ states
and the SOI parameters within the Hartree-Fock 
approximation (HFA).\cite{Cowan81}
The values are listed in Table \ref{table1}.
It is known that the values for the isotropic part of 
the Slater integral $F^0$ are considerably reduced due to large screening
effects. We reduce the value by multiplying a factor 0.25.
On the other hand, the values for the anisotropic part of the Slater integrals
are not screened so much. We slightly reduce them by multiplying a factor 0.8.
We diagonalize the Hamiltonian, and obtain nine degenerate states 
belonging to the $J=4$ subspace as the lowest-energy state.
Note that these states of $J=4$ are slightly deviated from those of the LS 
coupling scheme with $L=5$ and $S=1$ due to the SOI.

\begin{table}[t]
\caption{\label{table1}
Slater integrals and spin-orbit interaction parameters
within the HFA (in units of eV).\cite{Cowan81} \\
\centerline{the $(3d)^{10}(5f)^2$ configuration}}
\begin{center}
\begin{tabular}{llll}
\hline
\hline
 $F^{k}(3d,3d)$    & $F^{k}(3d,5f)$  & $F^{k}(5f,5f)$ & $G^{k}(3d,5f)$  \\
\hline
 $F^0$ \hspace*{0.2cm} 177.7 & 
  &
 $F^0$ \hspace*{0.2cm} 18.84 & 
 \\
 $F^2$ \hspace*{0.2cm} 90.80 & 
 &
 $F^2$ \hspace*{0.2cm} 9.519 &
 \\
 $F^4$ \hspace*{0.2cm} 58.48 &  
 &
 $F^4$ \hspace*{0.2cm} 6.228 &
 \\
                             &
                             &
 $F^6$ \hspace*{0.2cm} 4.573 & 
                             \\  
\hline
$\zeta_{3d}=$ 74.611 & $\zeta_{5f}=$ 0.277  & \\
\hline
\hline
\end{tabular}
{\footnotesize \centerline{the $(3d)^{9}(5f)^3$ configuration} }
\end{center}
\begin{center}
\begin{tabular}{llll}
\hline
\hline
 $F^0$ \hspace*{0.2cm} 178.7 & 
 $F^0$ \hspace*{0.2cm} 28.08 & 
 $F^0$ \hspace*{0.2cm} 19.82 & 
 $G^1$ \hspace*{0.2cm} 2.011 \\
 $F^2$ \hspace*{0.2cm} 91.39 & 
 $F^2$ \hspace*{0.2cm} 2.570 & 
 $F^2$ \hspace*{0.2cm} 10.03 &
 $G^3$ \hspace*{0.2cm} 1.216 \\
 $F^4$ \hspace*{0.2cm} 58.87 &  
 $F^4$ \hspace*{0.2cm} 1.194 & 
 $F^4$ \hspace*{0.2cm} 6.574 &
 $G^5$ \hspace*{0.2cm} 0.850 \\
                             &
                             &
 $F^6$ \hspace*{0.2cm} 4.832 & 
                             \\  
\hline
$\zeta_{3d}=$ 75.619 & $\zeta_{5f}=$ 0.318  & \\
\hline
\hline
\end{tabular}
\end{center}
\end{table}

It is known that the crystal structure of URu$_{2}$Si$_{2}$ is 
the body-centered tetragonal (bct)
ThCr$_{2}$Si$_{2}$ structure (I4/mmm) with lattice constants
$a=4.124 {\rm \AA}$ and $c=9.582 {\rm \AA}$
at $T=4.2$ K (Fig. \ref{fig.struc}).\cite{Mason90}
Under the bct symmetry, the degeneracy in the $J=4$
subspace is lifted by the CEF Hamiltonian, which may be expressed as
\begin{equation}
H_{\rm cry} = \sum_k \sum_q B_k^q O_k^q,
\label{eq.crys}
\end{equation}
where $O_k^q$'s denote the Stevens operator equivalence.
The CEF parameters $B_k^q$'s take non-zero values for $(k,q)$$=$$(2,0)$, 
$(4,0)$, $(4,4)$, $(6,0)$ and $(6,4)$.
The eigenstates of the Hamiltonian split into five singlets
and two doublets as listed in Table \ref{table0}, where
the $\epsilon$, $\gamma$, $\alpha$ depend on the CEF parameters.
At present, it is difficult to determine the CEF parameters
in a convincing way.
In the following, we discuss two prevailing CEF level schemes.

\begin{table}[t]
\caption{\label{table0}Eigenstates of the CEF Hamiltonian eq. (\ref{eq.crys})
with $|M \rangle$ denotes the state of $J_z=M$.}
\begin{center}
\begin{tabular}{ll}
\hline
\hline
$\Gamma_{t 1}^{(1)}$ & $\epsilon( | + 4 \rangle + | - 4 \rangle )
                        + \gamma | 0 \rangle$ \\ 
$\Gamma_{t 1}^{(2)}$ & $\frac{\gamma}{\sqrt{2}}
                        \left( | + 4 \rangle + | - 4 \rangle \right)
                        - \sqrt{2} \epsilon | 0 \rangle$ \\ 
$\Gamma_{t 2}$ & $\frac{1}{\sqrt{2}}
                        \left( | + 4 \rangle - | - 4 \rangle \right)$ \\
$\Gamma_{t 3}$ & $\frac{1}{\sqrt{2}}
                        \left( | + 2 \rangle + | - 2 \rangle \right)$ \\
$\Gamma_{t 4}$ & $\frac{1}{\sqrt{2}}
                        \left( | + 2 \rangle - | - 2 \rangle \right)$ \\
\hline
$\Gamma_{t 5}^{(1)}$ & $\cos \alpha | \pm 3 \rangle 
                        + \sin \alpha | \mp 1 \rangle$ \\ 
$\Gamma_{t 5}^{(2)}$ & $\sin \alpha | \pm 3 \rangle 
                        - \cos \alpha | \mp 1 \rangle$ \\ 
\hline
\hline
\end{tabular}
\end{center}
\end{table}

\subsubsection{Singlet scheme}
Analyzing the temperature dependences of magnetic susceptibility 
and specific heat, Santini and Amoretti\cite{Santini94}
have proposed a CEF level scheme 
that the lowest three states become $\Gamma_{t 3}$ (or $\Gamma_{t 4}$), 
$\Gamma_{t 1}^{(1)}$ and $\Gamma_{t 2}$ in order, with energy separation 
3.8 meV, 9.6 meV from the lowest state, respectively, and that
the other states have energies more than $\sim 40$ meV higher than the above.
This scheme has been considered for bringing about 
the AFQ ordering at low temperatures.
It successfully reproduced the experimental magnetic susceptibility
and specific heat in a wide temperature region.\cite{Santini94}

Within the states $\Gamma_{t 3}$, $\Gamma_{t 1}^{(1)}$ and $\Gamma_{t 2}$,
the dipole operator $J_z$ and the quadrupole operators
$O_{x^2-y^2}$ $\equiv$ $\frac{\sqrt{3}}{2}$ $(J_x^2-J_y^2)$,
$O_{xy}$ $\equiv$ $\frac{\sqrt{3}}{2}$ $(J_xJ_y+J_yJ_x)$ are represented as
\begin{align}
J_{z} &= \frac{8 \epsilon}{\sqrt{2}} \left( \begin{array}{ccc}
       0 & 0 & 0 \\
       0 & 0 & 1 \\
       0 & 1 & 0 \\
     \end{array} \right), \label{eq.Jz33}\\
O_{x^2-y^2} &= 
\sqrt{6} ( \sqrt{7} \epsilon + 3 \sqrt{5} \gamma )
\left( \begin{array}{ccc}
       0 & 1 & 0 \\
       1 & 0 & 0 \\
       0 & 0 & 0 \\
     \end{array} \right), \\
O_{xy} &= \sqrt{7}  \left( \begin{array}{ccc}
       0 & 0 & {\rm i} \\
       0 & 0 & 0 \\
      -{\rm i} & 0 & 0 \\
     \end{array} \right).
\end{align}
Another choice of the states $\Gamma_{t 4}$, $\Gamma_{t 1}^{(1)}$, 
$\Gamma_{t 2}$ makes the matrix of $O_{x^2-y^2}$ and that of $O_{xy}$ 
interchanged.
The $\epsilon$ and $\gamma$ are estimated as $-0.6775$ and $0.2864$,
according to Santini and Amoretti's analysis.\cite{Santini94}
For describing the AFQ phase, we add a term $\lambda Q\langle Q\rangle$
on the basis of the mean-field approximation to the multipolar interaction.
Here $\langle Q\rangle$ is 
the staggered moment of one of the quadrupole $O_{x^2-y^2}$ or $O_{xy}$
and $\lambda=0.185$ meV is a mean-field coefficient.\cite{Santini94,com2}  
The self-consistent solution gives us the AFQ ordering at low temperatures;
$|\langle Q\rangle| = 12.1$ at zero temperature.
The above expressions indicate that 
the $O_{x^2-y^2}$ (or $O_{xy}$) ordering is most realized 
at low temperatures through the off-diagonal element between
the lowest and the first excited states.

For describing the AFM phase, we now introduce a molecular field term 
$\lambda' J_z\langle J_z\rangle$, where $\langle J_z\rangle$ is 
the staggered dipole moment.
Since $J_z$ has no matrix element to the lowest energy level 
(see eq.~(\ref{eq.Jz33})), 
the molecular field energy has to exceed
the crystal field energy for giving rise to the AFM phase
at low temperature. For $\lambda'>0.468$ meV, 
a self-consistent solution exists as
$|\langle J_z\rangle|>3.490$ with
$\mu/\mu_B=|\langle L_z+2S_z\rangle|=2.887$.
For instance, when $\lambda'=0.5$ meV,
$|\langle J_z\rangle|=3.540$, and the staggered moment becomes 
$\mu=2.925 \mu_{\rm B}$.
Therefore the AFM ordering can be brought about
only with assuming a large intersite U-U coupling.
This staggered moment is much larger than 
the experimental value $\mu \simeq 0.4 \mu_{\rm B}$ 
in the pressure range between $p=1.5$ and $2.8$ GPa.
\cite{Amitsuka99,Amitsuka00}  

\subsubsection{Doublet scheme}
Ohkawa and Shimizu\cite{Ohkawa99} have proposed another model
that the lowest energy levels belong to the doublet 
$\Gamma_{t 5}^{(1)}$ or $\Gamma_{t 5}^{(2)}$.
This doublet scheme naturally gives rise to 
the AFQ and AFM orderings under moderate intersite U-U couplings.
Also it is consistent with the study of the dilute U$_x$Th$_{1-x}$Ru$_2$Si$_2$ 
alloys ($0 \leq x \leq 0.07$) by Amitsuka and Sakakibara,
which indicates that the lowest CEF state
of $5f$ electrons is magnetically degenerate
due to the absence of saturation in the susceptibility of 
U contribution down to 100 mK.\cite{Amitsuka94}

Choosing $\Gamma_{t 5}^{(1)}$ as a lowest doublet, we explicitly write 
\begin{eqnarray}
| \uparrow \rangle &=& \cos \alpha | +3 \rangle + \sin \alpha |-1 \rangle,
\label{eq.1} \\
| \downarrow \rangle &=& \cos \alpha |-3 \rangle + \sin \alpha |+1 \rangle,
\label{eq.2}
\end{eqnarray}
with $|M \rangle$ denoting the state of $J_z=M$.
Operators $J_z$, $O_{x^2-y^2}$ and $O_{xy}$ are represented
within the subspace as
\begin{align}
J_z &= (3 \cos^2 \alpha - \sin^2 \alpha ) 
\left( \begin{array}{cc}
            1 & 0  \\
            0 & -1 \\
       \end{array} \right), \label{eq.3.1} \\
O_{x^2-y^2} &= \sqrt{3} \sin \alpha (5 \sin \alpha + 3 \sqrt{7} \cos \alpha ) 
\left( \begin{array}{cc}
            0 & 1 \\
            1 & 0 \\
       \end{array} \right), \label{eq.3.2} \\
O_{xy} &= \sqrt{3} \sin \alpha (-5 \sin \alpha + 3 \sqrt{7} \cos \alpha ) 
\left( \begin{array}{cc}
            0 & -{\rm i} \\
            {\rm i} & 0 \\
       \end{array} \right). \label{eq.3.3} 
\end{align}
The intersite interaction induces the AFQ ordering or the AFM ordering.

For the AFQ ordering, the ground state is given by
an alternating array of the states $|+\rangle$ and $|-\rangle$, where
\begin{equation}
| \pm \rangle = \sqrt{\frac{1}{2}} \left[
       | \uparrow \rangle \pm | \downarrow \rangle \right], 
\end{equation}
for the $O_{x^2-y^2}$ ordering, and
\begin{equation}
| \pm \rangle = \sqrt{\frac{1}{2}} \left[
       | \uparrow \rangle \pm {\rm i} | \downarrow \rangle \right],
\end{equation}
for the $O_{xy}$ ordering.

For the AFM ordering, the ground state is just
an alternating array of $|\uparrow\rangle$ and $|\downarrow\rangle$. 
The staggered magnetic moment is evaluated from 
\begin{equation}
 \mu/\mu_B=\cos^2\alpha|\langle +3|L_z+2S_z|+3\rangle|
  +\sin^2\alpha|\langle -1|L_z+2S_z|-1\rangle|.
\label{eq.staggered}
\end{equation}
Putting $\alpha  \simeq 52.48^{\circ}$
or $\alpha \simeq 68.95^{\circ}$,
we obtain $\mu \simeq 0.4 \mu_{\rm B}$, close to the experimental value.
\cite{Amitsuka99,Amitsuka00}  

\subsection{The intermediate state}

The $E_1$ transition causes a $3d$-core electron to an unoccupied $5f$ states,
leading to the $(3d)^{9}(5f)^3$-configuration.
The Slater integrals and the SOI parameters
are evaluated within the HFA, which
are also summarized in Table \ref{table1}.\cite{Cowan81}
The Coulomb interactions and the SOI of $3d$ and $5f$ electrons are represented
within 364 $\times (2j_d+1)$ states in the $(3d)^{9}(5f)^3$-configuration
to construct the Hamiltonian,
and the intermediate state $|\Lambda\rangle$ is calculated
by diagonalizing the Hamiltonian.
Here $j_d=\frac{3}{2}$ and $\frac{5}{2}$ mean the total moment
of the $3d$-core electron from the $M_{\rm IV}$ and $M_{\rm V}$ edges,
respectively.
The CEF energy is so small in comparison with the multiplet energies
that it can be neglected in the intermediate state.
Then the energy levels are degenerate with respect to the magnetic quantum
number $M$ within the space of the total angular momentum J.

\section{\label{sect.4}Calculated Results} 

\subsection{Absorption coefficient}
We discuss the absorption coefficient $A(\omega)$
at U $M_{\rm IV}$ and $M_{\rm V}$ edges.
Within the $E_1$ transition, it is expressed as
\begin{equation}
 A(\omega) \propto
  \sum_j\sum_{\Lambda}\sum_{\alpha}|\langle\Lambda|x_{\alpha}(j)|\psi_0\rangle
  |^2\frac{\Gamma}{(\hbar\omega-E_{\Lambda})^2+\Gamma^2}.
\label{eq.absorption}
\end{equation}
We calculate $A(\omega)$ by assuming the AFQ ordering or the AFM ordering
at zero temperature. The final state of photoabsorption is the same as 
the intermediate state of the RXS explained in the preceding section.
The $\Gamma$ is assumed to be 0.5 eV.
Figure 3 shows the calculated result in comparison with the 
experiment.\cite{Yaouanc98}
The core-level energy is adjusted such that the calculated peak position
coincides with that of the experimental peak position.
The obtained spectra are independent of the choice of
the ordered phase, which can be verified by means of the results
obtained in Appendix.
The broad peak is due to multiplet structure of the final state.

\begin{figure}[t]
\begin{center}
\includegraphics[width=8.0cm]{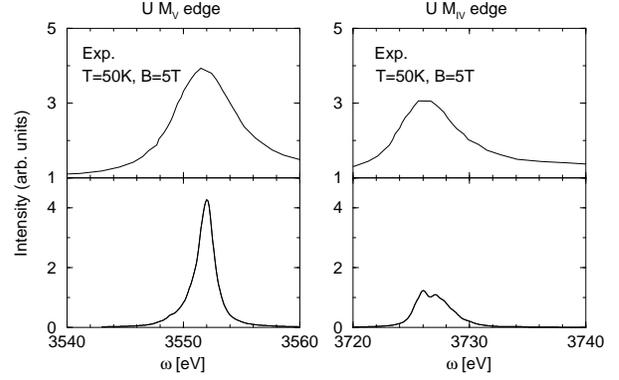}%
\end{center}
\caption{\label{fig.abs.p52}Absorption coefficient
as a function of the photon energy
near U $M_{\rm IV}$ and U $M_{\rm V}$ edges.
The upper panel is the experimental result (Ref. 48)
and the lower panel is the calculated result.}
\end{figure}

\subsection{RXS Spectra}

\subsubsection{AFQ phase}

Assuming the AFQ phase in the initial state and using the intermediate state
in the preceding section,
we calculate the RXS spectra at U $M_{\rm IV}$ and $M_{\rm V}$ edges
from eq.~(\ref{eq.dipole}). The scattering amplitude has a form
\begin{subequations}
\begin{align}
 M({\bf G},\omega) & \propto 
\xi'(\omega)   \left( \begin{array}{ccc}
       1 & 0  & 0 \\
       0 &-1 & 0 \\
       0 & 0 & 0 \\
    \end{array} \right), \quad \mbox{for} \ O_{x^2-y^2} \ \mbox{ordering}, 
\label{eq.afq.amp.1} \\
  & \propto \xi'(\omega)
   \left( \begin{array}{ccc}
       0 & 1 & 0 \\
       1 & 0 & 0 \\
       0 & 0 & 0 \\
    \end{array} \right), \quad
    \mbox{for} \ O_{xy} \ \mbox{ordering}. 
\label{eq.afq.amp.2}
\end{align}
\end{subequations}
Combining these expressions with the geometrical 
factors in eq.~(\ref{eq.rxs.intensity}),
we obtain the intensities for ${\bf G}=(00\ell)$
\begin{equation}
I_{\sigma \rightarrow \mu'} \propto \left\{
\begin{array}{lc} 
       \cos^2 (2 \psi), & \mu' = \sigma' \\
       \sin^2 (2 \psi) \sin^2 \theta, & \mu' = \pi' \\
\end{array} \right. ,
\label{eq.x2y2}
\end{equation}
for the $O_{x^2-y^2}$ ordering, and 
\begin{equation}
I_{\sigma \rightarrow \mu'} \propto \left\{
\begin{array}{lc} 
       \sin^2 (2 \psi), & \mu' = \sigma' \\
       \cos^2 (2 \psi) \sin^2 \theta, & \mu' = \pi' \\
\end{array} \right. ,
\label{eq.xy}
\end{equation}
for the $O_{xy}$ ordering.

\begin{figure}[t]
\begin{center}
\includegraphics[width=8.0cm]{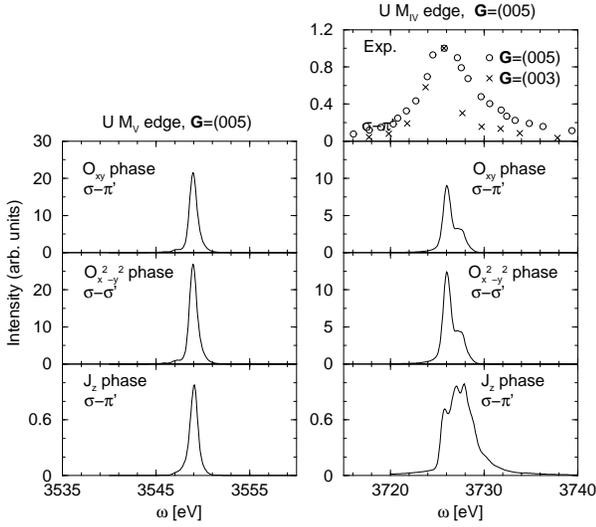}%
\end{center}
\caption{\label{fig.m4.m5.p52}The RXS spectra as a function of
the photon energy at ${\bf G}=(0 0 5)$ with $\psi=0$.
Left and right panels show the results around
U $M_{\rm V}$ edge and $M_{\rm IV}$ edge, respectively.
Top right panel shows the experimental result.\cite{Isaacs90,Bernhoeft03.1}
The rests are calculated results
corresponding to those in the $J_{z}$, $O_{x^2-y^2}$ and $O_{xy}$
phases from bottom to top, respectively,
for both sides of the panels.}
\end{figure}

We find that the spectral shapes as a function of energy are independent
of the assumed level schemes, although the intensities are quite different
between the schemes.\cite{com1}
Figure \ref{fig.m4.m5.p52} shows the RXS spectra at ${\bf G}=(005)$ with
$\psi=0$ near U $M_{\rm IV}$ and $M_{\rm V}$ edges,
as a function of photon energy.
For the $M_{\rm IV}$ edge, the spectra have a width of $\sim 5$ eV 
due to the multiplet structure in the intermediate state;
a large peak is at $\hbar\omega=3.726$ keV. and 
a shoulder is at $\sim 2$ eV higher 
than the peak. The calculated width looks somewhat narrower than
the experimental one.\cite{Isaacs90,Bernhoeft03.1}
For the $M_{\rm V}$, a narrower peak is found at 
$\hbar\omega=3.549$ keV.

As clear from eqs.~(\ref{eq.x2y2}) and (\ref{eq.xy}), the RXS intensity depends 
considerably on the azimuthal angle. Figure 5 demonstrates this dependence.
This is in contrast with the RXS intensity in the AFM state,
where it shows no $\psi$ dependence as shown below.

\begin{figure}[t]
\begin{center}
\includegraphics[width=8.0cm]{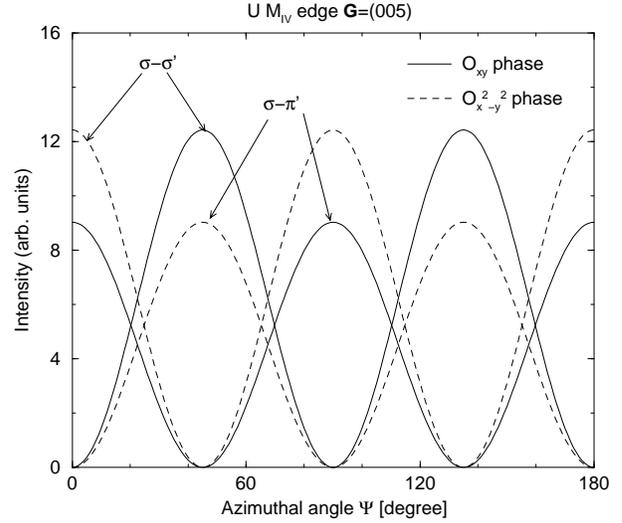}
\end{center}
\caption{\label{fig.azim.p52}Azimuthal angle dependence of the 
RXS peak intensities at ${\bf G}=(0 0 5)$ near
U $M_{\rm IV}$ edge in the AFQ phases.
The solid and broken lines represent the results obtained
in the $O_{xy}$  and $O_{x^2-y^2}$ orderings
within the doublet scheme, respectively.
The mixing angle is fixed to being $\alpha \simeq 52.48^{\circ}$
and both $\sigma-\sigma'$ and $\sigma-\pi'$ channels are shown.}
\end{figure}

\subsubsection{AFM phase}

We also calculate the RXS spectra, assuming the AFM phase in the initial state.
The scattering amplitude $M({\bf G},\omega)$ has a form
\begin{equation}
 M({\bf G},\omega) \propto \left( \begin{array}{ccc}
       0 & \xi(\omega) & 0 \\
       -\xi(\omega) & 0 & 0 \\
       0 & 0 & 0 \\
    \end{array} \right).
\label{eq.afm.amp}
\end{equation}
Combining this with the geometrical factors in eq. (\ref{eq.rxs.intensity}), 
we can verify that the RXS intensities show no signal 
in the $\sigma-\sigma'$ channel while
those in the $\sigma-\pi'$ channel 
remain constant with varying azimuthal angle $\psi$.
This azimuthal angle dependence is valid as long as
the scattering vector is in the form of $(00\ell)$.

As the same as the case of AFQ state, we find that the spectral shapes
as a function of photon energy are independent of the assumed level schemes.
\cite{com1}
The RXS spectra at ${\bf G}=(005)$ near U $M_{\rm IV}$ and $M_{\rm V}$ edges
are shown in Fig.~\ref{fig.m4.m5.p52}.
The spectral peak is located at  position
$\sim 2$ eV higher than that in the AFQ state, and the shoulder is seen 
at the lower energy side.
This spectral shape is clearly different from that in the AFQ state. 
Unfortunately, the experimental spectra are not clear enough to
distinguish the difference.

\section{\label{sect.5} Concluding Remarks}

We have analyzed the RXS spectra near U $M_{\rm IV}$ and $M_{\rm V}$ edges
on the basis of a localized CEF model.
By employing two level schemes, the singlet model of Santini and 
Amoretti and the doublet model of Ohkawa and Shimizu, we have considered
the AFQ phase as a candidate of the HO phase and the AFM phase 
as the one realized under high pressure $p>1.5$ GPa.
We have calculated the RXS spectra by
taking into account the full Coulomb and spin-orbit interaction
within the initial $(3d)^{10}(5f)^2$ and intermediate $(3d)^{9}(5f)^3$ 
configurations.

We have found that the spectral shapes as a function of photon energy
are the same in both schemes. This looks unexpected, because  
the wave functions of the initial state are different between the schemes.
Our numerical finding can be proved 
if the crystal field energy is negligible in the intermediate state,
as given in Appendix.

The obtained spectra show strong differences between
U $M_{\rm IV}$ and $M_{\rm V}$ edges, the former exhibits
the broader and more complicated (multi-peak structure) profile 
than the latter as a function of photon energy. 
Also the spectral shapes are found quite different between the AFQ phase
and the AFM phase.
These differences as well as the photon polarization dependence may provide
useful guidance to elucidate the symmetrical aspects of the actual system
when the RXS experiment will be performed in high pressure phase.

The spectral shape calculated in the AFQ phase as a function of photon energy 
seems consistent with the experimental one at ambient pressure.
On the other hand, the spectra have to depend on the azimuthal angle,
as shown in Fig.~5.
Quite recently, Bernhoeft \textit{et al}. studied
the azimuthal angle dependence of the peak intensity on the
(005) magnetic spot in URu$_{2}$Si$_{2}$,\cite{Bernhoeft03.1}
having claimed that the spectra show no azimuthal angle dependence.
Mixture of phases of the $O_{x^2-y^2}$ ordering and the $O_{xy}$ ordering
may make the spectra less depending on the azimuthal angle.
Otherwise this may rule out the AFQ type ordering as a possible
HO candidate.

Finally, URu$_2$Si$_2$ is considered to be located
at the borderline between a localized and an itinerant description.
The CEF framework adopted in this work is to be seen
only as a first approximation for the comprehensive understanding
of the complex physics of this compound. The effort to this direction 
will be the next study. 


\section*{Acknowledgments}
We would like to thank H. Amitsuka and M. Yokoyama for valuable discussions.
J.I. thanks Max-Planck Institute for the Physics of Complex Systems
for hospitality during his stay, where this work was completed.
This work was partially supported by a Grant-in-Aid for Scientific Research 
from the Ministry of Education, Science, Sports and Culture.


\appendix

\section{Properties of the Scattering Amplitude}

In this appendix, we derive general expressions of the matrix elements
of the scattering amplitude in the $E_1$ process, assuming that
the intermediate state maintains rotational symmetry.
This assumption seems reasonable when the CEF energy is negligible 
compared with the magnitude of the multiplet  splitting
in the intermediate state. Such conditions would be satisfied 
in many rare earth metal compounds. We show that
the obtained expressions lead us to a useful limit on 
the shape of the RXS spectra as a function of the photon energy;
it depends only on the type of the order parameter.
We will not discuss the case of the $E_2$ transition,
since any useful limits on the spectral shape have not been derived
from a similar discussion applied to this case.
Although a similar discussion
had been done by Lovesey and Balcar,\cite{Lovesey96}
we treat rigorously
the energy denominator
appeared in the eq. (\ref{eq.dipole}),
which is set aside in ref.51,
in order to investigate the energy dependence
of the RXS spectra.

\subsection{General results}

First, we rewrite the scattering tensor in eq.  (\ref{eq.dipole})
as
\begin{align}
M_{\alpha \alpha'}({\bf G},\omega) &= \frac{1}{\sqrt{N}}
\sum_{j} {\rm e}^{-{\rm i} {\bf G}\cdot {\bf r}_j} 
\tilde{M}_{\alpha \alpha';j}(\omega), \\
\tilde{M}_{\alpha \alpha';j}(\omega)
&\equiv 
\sum_{\Lambda} E(\omega,\Lambda)
  \langle\psi_0|x_\alpha(j)|\Lambda\rangle
  \langle \Lambda|x_{\alpha'}(j)|\psi_0\rangle, \nonumber \\
\label{eq.mdef.1} \\
E(\omega,\Lambda) &\equiv \frac{1}
       {\hbar\omega-(E_{\Lambda}-E_0)+i\Gamma}.
\end{align}
Note that 
contributions from the sites belonging to the same sublattice are
the same.  We omit the subscript $j$ in the following.
The initial state can be expanded by a linear combination
of $| J, m \rangle$ as
\begin{equation}
| \psi_0 \rangle = \sum_{m} c(m) | J, m \rangle,
\end{equation}
within a ground multiplet of $J$ with $m$ being the eigenvalue
of $J_z$.
If the rotational invariance is preserved in the intermediate 
state, the eigenstates $|\Lambda \rangle$ are those of ${\rm J}^2$
and $J_z$ simultaneously. (Of course the intermediate state is occupied
by a core hole.)   
Notice that for a given value of $J$, there exist more than one
multiplets having the same $J$ value but having the different energy.
Therefore we introduce variable $i$ telling from the multiplets 
with the same value of $J$, and write the intermediate state as
$|\Lambda \rangle = |J', m' , i\rangle$.
Then eq. (\ref{eq.mdef.1}) is written as
\begin{align}
 \tilde{M}_{\alpha \alpha'}(\omega)
&=
 \sum_{m,m'} c^{\star}(m) c(m') 
\hat{M}_{\alpha \alpha'}^{m,m'}(\omega), \\
 \hat{M}_{\alpha \alpha'}^{m,m'}(\omega)
&\equiv
\sum_{J'} \sum_{i=1}^{N_{J'}} \sum_{M=-J'}^{J'} E_i(\omega,J')
\nonumber \\
&\times
  \langle J, m|x_\alpha|J', M, i \rangle
  \langle J', M, i |x_{\alpha'}| J,m'\rangle, 
\end{align}
where $E_i(\omega,J')$ represents $E(\omega,\Lambda)$ 
in eq. (\ref{eq.mdef.1}).
The number of the multiplets having the value $J$ is denoted by $N_J$.
The selection rule for the $E_1$ process confines the range of the summation
over $J'$ to $J'=J, J \pm 1$.
The evaluation of the matrix element of the type
$\langle J, m|x_\alpha|J', m' \rangle$ is yielded by utilizing the
Wigner-Eckart theorem for a vector operator.\cite{Condon35}
We write down several results derived from 
$\hat{M}_{\alpha \alpha'}^{m,m'}(\omega)$
in the following.

First, 
$\hat{M}^{m,m '}(\omega) \neq {\bf 0}$ only when $|m -m '| \leq 2$.
It is obvious because of the nature of the dipole operators.
Second, for $m'=m$, 
\begin{align}
\hat{M}^{m,m} (\omega)&=
\left( \begin{array}{ccc}
 a_{m}(\omega) & m \alpha (\omega)  & 0 \\
-m \alpha(\omega)   & a_{m}(\omega) & 0 \\
 0 & 0 & b_{m}(\omega) \\
\end{array} \right), \\
a_m(\omega) &=  [ J(J+1) + m^2] F_{J-1}(\omega) \nonumber \\
&+ [ J(J+1) -m^2 ] F_{J}(\omega) \nonumber \\
&+ [ J(J+3) +m^2 +2 ] F_{J+1}(\omega), \\
b_m(\omega) &=  (J^2-m^2 ) F_{J-1}(\omega) + m^2 F_{J}(\omega)
\nonumber \\
&+ [ J(J+2) -m^2 +1 ] F_{J+1}(\omega), \\
\alpha(\omega) &=
 - (2J+1) F_{J-1}(\omega) \nonumber \\
&- F_{J}(\omega) + (2J+3) F_{J+1}(\omega), \\
F_{J'}(\omega) &\equiv |(J||V_{1}||J')|^2
 \sum_{i=1}^{N_{J'}} E_i(\omega,J'), 
\end{align}
where $(J||V_{1}||J')$ denotes the reduced matrix element
of the set of irreducible tensor operator of the first rank.
Note that the diagonal elements satisfy the relations 
\begin{equation}
 a_{m}(\omega) = a_{-m}(\omega), \quad b_{m}(\omega) = b_{-m}(\omega).
\end{equation}
One remarkable feature is that the off-diagonal elements are just 
proportional to $m$, completely factorized by the same 
function $\alpha(\omega)$.
Another important result is that the trace is independent of $m$
\begin{align}
{\rm Tr} [M^{m,m}(\omega) ]
&= 2 a_m(\omega)+b_m(\omega) 
\nonumber \\
&= J(2J+1)F_{J-1}(\omega) + J(J+1) F_{J}(\omega) \nonumber \\
&+ [ J(2J+5)+1 ] F_{J+1}(\omega),
\end{align}
where ${\rm Tr} X$ is the trace of a quantity $X$.

Third, when $|m' - m|=2$, we can derive the following results.
\begin{align}
 \hat{M}^{m,m+2}  &= a_{m}'' \alpha''(\omega)
\left( \begin{array}{ccc}
    1      &  {\rm i} & 0 \\
 {\rm i}   & - 1    & 0 \\
 0 & 0 & 0 \\
\end{array} \right), \\
 \hat{M}^{m+2,m}  &= a_{m}'' \alpha''(\omega)
\left( \begin{array}{ccc}
     1     & - {\rm i} & 0 \\
 - {\rm i} & - 1    & 0 \\
 0 & 0 & 0 \\
\end{array} \right),  \\
 a_{m}''  &=  \sqrt{(J-m)(J+m+1)} \nonumber \\
          &\times  \sqrt{(J-m-1)(J+m+2)}, \\
\alpha''(\omega) &\equiv
  - F_{J-1}(\omega) + F_{J}(\omega) - F_{J+1}(\omega).
\end{align}
Here, we can see the energy dependence is factorized by the 
same function $\alpha''(\omega)$ for every $m$.
From these results, 
we can check $a_{-(m+2)}'' = a_m''$ holds, which
leads us to the following relation
\begin{equation}
\hat{M}^{m,m \pm 2}(\omega) = \hat{M}^{-(m \pm 2),-m}(\omega), 
\end{equation}
where either all the upper or all the lower signs are to be taken.

Finally, for another $m' \neq m$ case, that is, for $|m'-m|=1$, 
the scattering amplitudes take forms
\begin{align}
 \hat{M}^{m,m+1}(\omega) &= 
\left( \begin{array}{ccc}
 0  & 0 & a_{m}' (\omega)\\
 0  & 0 & {\rm i} a_{m}' (\omega)\\
 b_{m}'(\omega) & {\rm i} b_{m}'(\omega) & 0 \\
\end{array} \right), \\
 \hat{M}^{m+1,m} (\omega)&=
\left( \begin{array}{ccc}
 0  & 0 & b_{m}'(\omega) \\
 0  & 0 &-{\rm i} b_{m}'(\omega) \\
 a_{m}'(\omega) &-{\rm i} a_{m}'(\omega) & 0 \\
\end{array} \right). 
\end{align}
Actual forms of $a_{m}'(\omega)$ and $b_{m}'(\omega)$ can be obtained, but
are not listed here because they are irrelevant to our present interest.
We only notice that the matrix elements $a_m'(\omega)$ and $b_m'(\omega)$ 
cannot separate their arguments $\omega$ and $m$.

\subsection{Application to URu$_2$Si$_2$}

In the AFM phase, the relevant contributions to the RXS amplitude
stem from the terms $\hat{M}^{m,m}(\omega)-\hat{M}^{-m,-m}(\omega)$
$\propto m \alpha(\omega)$, which justifies the matrix form given by
eq. (\ref{eq.afm.amp}).  A detailed calculation reveals that the amplitude
is proportional to the expectation value $\langle J_z \rangle$.

In the AFQ phases, on the other hand,
the relevant contributions to the RXS amplitude
come from the terms $\hat{M}^{m,m+2}(\omega) \pm \hat{M}^{m+2,m}(\omega)$
$\propto a_{m}'' \alpha''(\omega)$ where a plus sign corresponds to
the $O_{x^2-y^2}$ phase while a minus sign corresponds to the $O_{xy}$ phase.
This also verifies the scattering amplitudes are given by
eqs. (\ref{eq.afq.amp.1}) and (\ref{eq.afq.amp.2}).
In these ordering phases, the energy dependence of the
scattering amplitude is factorized 
by a single function $\alpha(\omega)$ in the AFM phase or $\alpha''(\omega)$
in the AFQ phase.
This is the reason why the shape of the RXS spectra depends only 
on the type of the order parameter.
Note that this factorization is special for the $E_1$
transition in the present model.
If terms $\hat{M}^{m,m\pm 1}(\omega)$ become relevant for some
physical reason, the factorization breaks down due to the presence of
the elements $a_m'(\omega)$ and $b_m'(\omega)$.
We can see that the situation is more complicated for the $E_2$ transition.

Since the true nature of the hidden order phase is unknown,
it might be useful to  consider what happens if some 
octupolar phase is substantiated in URu$_2$Si$_2$.
Assuming two components of the octupolar ordering
$T_{xyz}$ and $T_{z}^{\beta}$,
we can calculate the RXS signal from the octupole phase.
For instance, we choose the initial states as these 
ordering phases
in the singlet scheme and apply the results obtained in this appendix.
The result says no RXS signal is detected in the 
antiferrooctupolar phase.


\begin{thebibliography}{99}
\bibitem{com1}
This unexpected property can be proved if 
the intermediate state has rotational symmetry. 
This condition is satisfied within a good approximation
if the crystal field energy is much smaller 
than the multiplet energy.
\bibitem{com2} 
The definitions of the components of the quadrupolar operators 
adopted in ref. 13 
are different from those of the present work by a factor $\sqrt{3}/2$.
Therefore, the value of the coupling constant $\lambda$ in ref. 13 
corresponds to $4 \lambda/3$ here.
\end{thebibliography}
\end{document}